\documentclass[aps,preprint,nofootinbib,floatfix]{revtex4}
\usepackage{xcolor}
\usepackage{graphicx,color,soul}
\usepackage[latin1]{inputenc}
\usepackage{amsmath,amssymb}
\usepackage{hyperref}
\usepackage{epstopdf}
\usepackage{ulem}

\newcommand{\lsim}{\mathrel{\mathop{\kern 0pt \rlap
  {\raise.2ex\hbox{$<$}}}
  \lower.9ex\hbox{\kern-.190em $\sim$}}}
\newcommand{\gsim}{\mathrel{\mathop{\kern 0pt \rlap
  {\raise.2ex\hbox{$>$}}}
  \lower.9ex\hbox{\kern-.190em $\sim$}}}

\newcommand{\feff}{\ensuremath{f_{\texttt{eff}}}}

\newcommand{\gev}{\ensuremath{\,\mathrm{GeV}}}
\newcommand{\tev}{\ensuremath{\,\mathrm{TeV}}}

\newcommand{\mk}{\ensuremath{\,\mathrm{mK}}}
\newcommand{\mchi}{\ensuremath{m_\chi}}
\newcommand{\sv}{\ensuremath{\langle\sigma v\rangle}}
\newcommand{\dngde}{\ensuremath{\frac{dN_\gamma}{dE_{\gamma}}}}
\newcommand{\dnede}{\ensuremath{\frac{dN_{e^+}}{dE_{e^+}}}}

\begin{document}
\title{The impact of EDGES 21-cm data on dark matter interactions}
\author{Kingman Cheung$^{a,b,c}$}
\author{Jui-Lin Kuo$^{a}$}
\author{Kin-Wang Ng$^{d,e}$}
\author{Yue-Lin Sming Tsai$^{d}$}

\affiliation{
$^{a}$Department of Physics, National Tsing Hua University, Hsinchu 30013, 
Taiwan\\
$^{b}$Physics Division, National Center for Theoretical Sciences, 
Hsinchu 30013, Taiwan\\
$^{c}$Division of Quantum Phases and Devices, School of Physics, 
Konkuk University, Seoul 143-701, Republic of Korea\\
$^{d}$Institute of Physics, Academia Sinica, Taipei 11529, Taiwan\\
$^{e}$Institute of Astronomy and Astrophysics, Academia Sinica, Taipei 11529, Taiwan 
}


\begin{abstract}
The recently announced results on the 21-cm absorption spectrum by the EDGES experiment
can place very stringent limits on dark matter annihilation cross sections. 
We properly take into account the heating energy released from dark matter annihilation from the radiation epoch to the 21-cm observation redshifts in the radiative transfer to compute the evolution of the gas temperature.
Our results show that the global 21-cm absorption profile is a powerful cosmological probe of the
dark matter interactions. For dark matter annihilating into electron-positron pairs, 
the EDGES results give a more stringent upper limit than the PLANCK result 
on the annihilation cross section at the lower dark matter mass region.
\end{abstract}

\maketitle

\section{Introduction \label{sec:introduction}}

The Big Bang theory is the most accepted theory for the beginning of the
Universe. The theory has been established by astrophysical and cosmological observations, especailly
the cosmic microwave background (CMB) in 90's~\cite{Boggess:1992xla}. 
The CMB, also called the ``afterglow'' of the Big Bang, was the earliest
light coming out from the soup of free electrons, protons, neutrons, and nuclei.
Only when the free electrons were caught by protons, neutrons, and 
nuclei to form neutral atoms, mostly hydrogen and helium,
the photons can shine through the matter and form the CMB.
It happened about 380,000 years after the Big Bang. 
Such footprints of the early Universe can tell us
a lot of information about the ingredients of the Universe.

Subsequently, the WMAP in the last decade pinned down the
valid parameters of the most popular cosmological model, namely, the
$\Lambda$CDM model~\cite{Hinshaw:2012aka}. The Planck collaboration~\cite{Ade:2015xua}
continued the mission of measuring more precisely the model parameters
in this decade, 
and particularly provided the measurements of the CMB polarization 
with improved precision.

After the recombination epoch, the Universe entered into the so-called
dark ages, containing mainly neutral hydrogen gas.
It is believed that during this period the dark matter (DM) first began to 
cluster to form halos that accrete normal matters.
The Universe then entered into the ``reionization'' epoch
when the first galaxy or star was being formed in the halo. 
The Gunn-Peterson test on
quasar absorption lines has revealed that the Universe has
been reionized at redshift $z\sim 7$~\cite{Gunn:1965hd}. More precise
measurements of the reionization bump in the CMB
$E$-mode polarization power spectrum have inferred an
optical depth integrated to the recombination, $\tau\sim 0.08$, and a
full reionization at $z\sim 9$~\cite{Ade:2015xua}. 
Nevertheless, how the reionization happened and when exactly it happened
are uncertain. It is generally believed that
radiation from first stars and/or galaxies ionize the neutral
gas beginning at $z\sim 15-20$~\cite{Pritchard:2011xb}.

A direct measurement of the reionization is made possible owing to the
hyperfine splitting of neutral hydrogen atoms, which emit or absorb
photons of wavelength at 21-cm in the rest frame dependent on the spin
temperature of the hydrogen gas, $T_s$. This spin temperature is
determined by the background temperatures of the CMB ($T_\gamma$) as
well as the surrounding thermal gas ($T_g$). Thus, measurements of 
redshifted 21-cm lines allow us to perform a tomographic study of the
reionization process. 
There have been a lot of observational efforts to
map the primeval hydrogen gas distributions; current experiments
include LOFAR~\cite{vanHaarlem:2013dsa}, MWA~\cite{Tingay:2012ps}, SKA~\cite{Maartens:2015mra}, and many
more~\cite{Pritchard:2011xb}. Recently, the EDGES experiment has
detected the global signal of 21-cm
absorption~\cite{Bowman:2018yin}, if confirmed, opening a new chapter of the 21-cm
cosmology.

EDGES has claimed that the measured amplitude of the 21-cm absorption
profile is more than a factor of two greater than the largest
predictions~\cite{Bowman:2018yin}. To explain this anomaly, one may have to
reduce the spin temperature $T_s$ 
by introducing novel photon-DM interactions to cool down the
gas~\cite{Barkana:2018lgd,Munoz:2018pzp,Barkana:2018qrx,Berlin:2018sjs,Fialkov:2018xre,Fraser:2018acy}. 
An alternative way is to add a strong radio background such that the
effective $T_\gamma$ is higher than the original CMB
temperature~\cite{Feng:2018rje,Ewall-Wice:2018bzf,Mirocha:2018cih,Pospelov:2018kdh}. 

In this work, instead of explaining the anomaly, 
we will use the EDGES result to constrain \textit{weakly-interacting massive} DM (WIMP) annihilation in the early Universe. 
In our approach we assume that the known or unknown sources 
more or less reproduce the EDGES signal. The DM annihilation that we are considering is
additional, though small, contributions to the EDGES signal.  
We can then use the uncertainty of the EDGES to constrain the DM interaction. 
Namely, the allowed parameter space of WIMP annihilation totally 
depends on the size of the EDGES experimental uncertainties.  

Generally speaking, as far as the WIMP annihilation is considered, 
one may naturally wonder if there is a sizable scattering cross section between DM and SM particles. 
Especially, they usually share the same couplings. 
If the size of the scattering cross section between DM and proton is considerable, 
the energy of gas can be taken away by the scattering process.  
However, a null signal of the traditional WIMP (having mass between $\gev$ to several $\tev$) 
has been obtained by current XENON1T~\cite{Aprile:2018dbl} and PandaX~\cite{Cui:2017nnn} 
and hence a severe limit has been reported. 
With such a low scattering cross section, 
we therefore do not need to consider the effect of gas cooling resulting from WIMP-baryon scatterings 
in the reionization epoch. 

If the DM annihilates into standard model (SM) particles such as quarks, leptons, and photons, 
they will heat up the gas and increase the gas temperature $T_{g}$. 
Since $T_s$ somewhat traces $T_{g}$, 
the DM annihilation will affect the evolution of $T_s$ and hence the $T_{21}$ signal probed by EDGES.
One may suspect that any limits on the DM interaction are entirely degenerate with the unknown mechanism that produces the EDGES absorption feature.
Firstly, the DM cooling of the gas has been severely constrained~\cite{Berlin:2018sjs,Fraser:2018acy}, it is unlikely that it can further compensate the heating by DM annihilation. 
Secondly, although increasing $T_\gamma$ also cancels the effect of
DM heating of the gas, the amount of radiation excess present at $z\sim
15-20$ would contribute to the radio radiation today and thus can be probed by
measurements of the radio background. 
This has been studied most recently in Ref.~\cite{Dowell:2018mdb}, where they have found support for
the existence of a strong diffuse radio background and suggested that additional radio
data would help understanding the Galactic foregrounds so as to constrain
the excess $T_\gamma$.
Thus, the DM annihilation rate would be excluded as $T_{s}$ is raised to a value beyond the experimental 
uncertainty of the EDGES absorption profile. 
Even though the EDGES signal turns out to be false or rectified to a predicted level, it is indeed
its experimental sensitivity that can be used as a gauge for constraining DM annihilation.

\section{Methodology \label{sec:method}}

EDGES has recently measured an absorption feature for 21-cm emission~\cite{Bowman:2018yin}.
At the redshift $z=17.2$, the temperature $T_{21}$ at $99\%$ 
confidence level (C.L.) is reported by 
\begin{eqnarray}
T_{21}^{\texttt{EDGES}}=-500^{+200}_{-500} \mk,   
\end{eqnarray}
where the errors $^{+200}_{-500}\mk$ present the systematic uncertainties.

On the other hand, the theoretical prediction is given by    
\begin{eqnarray}
T_{21}(z)\simeq 23 {\rm mK} \left[ 1- \frac{T_{\gamma}(z)}{T_s(z)}
 \right]
\left(\frac{\Omega_b h^2}{0.02} \right)
\left(\frac{0.15}{\Omega_m h^2} \right)
\sqrt{\frac{1+z}{10}} x_{HI}, 
\label{eq:t21}
\end{eqnarray}
where $\Omega_b h^2$ and $\Omega_m h^2$ are the relic densities of baryon and matter, respectively.
The number fraction of neutral hydrogen $x_{HI}$ is approximately equal to $1-x_e$,
where $x_e$ is the ionization fraction. 
The photon temperature $T_{\gamma}(z)$ can be the same as the CMB temperature, $T_{CMB}=2.7(1+z)$K, and 
the spin temperature $T_{s}(z)$ controls the atomic excitation between the $s=0$ and $s=1$ states of the neutral hydrogen. The 
value
of $T_{s}(z)$ lies between the gas temperature and the CMB temperature. 
Therefore, the precise value of $T_{s}(z)$ would be sensitive to the gas heated by DM annihilation.

In the early Universe, DM annihilates or decays into SM particles and 
ultimately produces some amount of energetic electrons/positrons ($e^-/e^+$)
and gamma rays ($\gamma$). Those of high-energy $e^{\pm}$ and $\gamma$ can 
ionize, heat, and excite the hydrogen atoms.  
We follow the formalism and methodology developed in 
Ref.~\cite{Kanzaki:2008qb,Kanzaki:2009hf,Kawasaki:2015peu},
which does not rely on any assumption on
the energy fractions $f_{\rm{eff}}$~\cite{Slatyer:2012yq,Slatyer:2015kla} 
as used in the literature related to CMB constraints on DM annihilation.
The DM contribution to the ionization fraction of hydrogen atom ($x_e$) via 
DM annihilation is 
\begin{eqnarray}
	\label{eq:dxdz}
	-\left[\frac{dx_e}{dz}\right]_{\rm DM} = \sum_\texttt{ch} \int_z \frac{dz'}{H(z')(1+z')}
	\frac{n_\chi^2(z') }{2n_H(z')} \sv \mathcal{B}(z')\texttt{BR}_\texttt{ch}
	\frac{m_\chi}{E_{\rm Ry}} \frac{d\chi_i(\texttt{ch},m_\chi,z',z)}{dz}, 
\end{eqnarray}  
and the gas temperature $T_g$ is modified as 
\begin{eqnarray}
	\label{eq:dTdz}	
	-\left[ \frac{dT_g}{dz}\right]_{\rm DM} = \sum_\texttt{ch} \int_z \frac{dz'}{H(z')(1+z')}
	\frac{n_\chi^2(z')}{3n_H(z')}\sv \mathcal{B}(z')\texttt{BR}_\texttt{ch}
	m_\chi \frac{d\chi_h(\texttt{ch},m_\chi,z',z)}{dz},
\end{eqnarray}  
where the subscript \texttt{'ch'} represents the 
DM annihilation or decay channels with 
the branching ratio $\texttt{BR}_\texttt{ch}$ and 
$i$ or $h$ refers to ionization or heating. 
%
%
Note that in the above expressions the energy is injected by 
DM annihilation at redshift $z'$ while absorbed by the neutral 
hydrogen at redshift $z$.
We define the parameters: the Rydberg energy $E_{\rm Ry}=13.6$\,eV, the DM mass $\mchi$, 
the DM annihilation cross section $\sv$, the DM number density $n_\chi$, 
and the number density of hydrogen atoms $n_H$. 
The fraction of injected energy for a given $\mchi$ and 
an annihilation channel \texttt{'ch'} can 
be obtained by integrating out the energy $E$, 
\begin{eqnarray}
	&&\frac{d\chi_{i,h}(\texttt{ch},m_\chi,z',z)}{dz}= \nonumber\\
	&&\int dE \frac{E}{m_\chi}\left[ 
		2\frac{dN_{e}(\texttt{ch},\mchi)}{dE} \frac{d\chi_{i,h}^{(e)}(E,z',z)}{dz} + 
		\frac{dN_{\gamma}(\texttt{ch},\mchi)}{dE} \frac{d\chi_{i,h}^{(\gamma)}(E,z',z)}{dz} 
	\right],
	\label{eq:dchidz}
\end{eqnarray}  
where the injected energy fractions for
electron and photon are, respectively,
\begin{align}
	 \frac{d\chi_{i,h}^{(e)}(E,z',z)}{dz}  {\rm ~~and~~}  \frac{d\chi_{i,h}^{(\gamma)}(E,z',z)}{dz},
\end{align}
given by Ref.~\cite{Kanzaki:2008qb,Kanzaki:2009hf,Kawasaki:2015peu}.
The energy spectrum per DM annihilation at the source, 
$\frac{dN_{e}(\texttt{ch},\mchi)}{dE}$ for electrons 
and $\frac{dN_{\gamma}(\texttt{ch},\mchi)}{dE}$ for photons, 
are calculated by \texttt{LikeDM}~\cite{Huang:2016pxg} 
by using the tables from \texttt{PPPC4}~\cite{Cirelli:2010xx}.
%
Our calculation of gas temperature evolution is similar to those in
Refs.~\cite{Chen:2003gz,Slatyer:2009yq,Evoli:2012zz}, 
except that we are using a set of tables for the transfer
functions derived in Ref.~\cite{Kanzaki:2008qb,Kanzaki:2009hf,Kawasaki:2015peu}, 
but Refs.~\cite{Chen:2003gz,Slatyer:2009yq,Evoli:2012zz} used the publicly
available results of $\feff$ in Ref.~\cite{Slatyer:2015kla}.

\begin{figure}[!htb]
\includegraphics[width=0.7\textwidth]{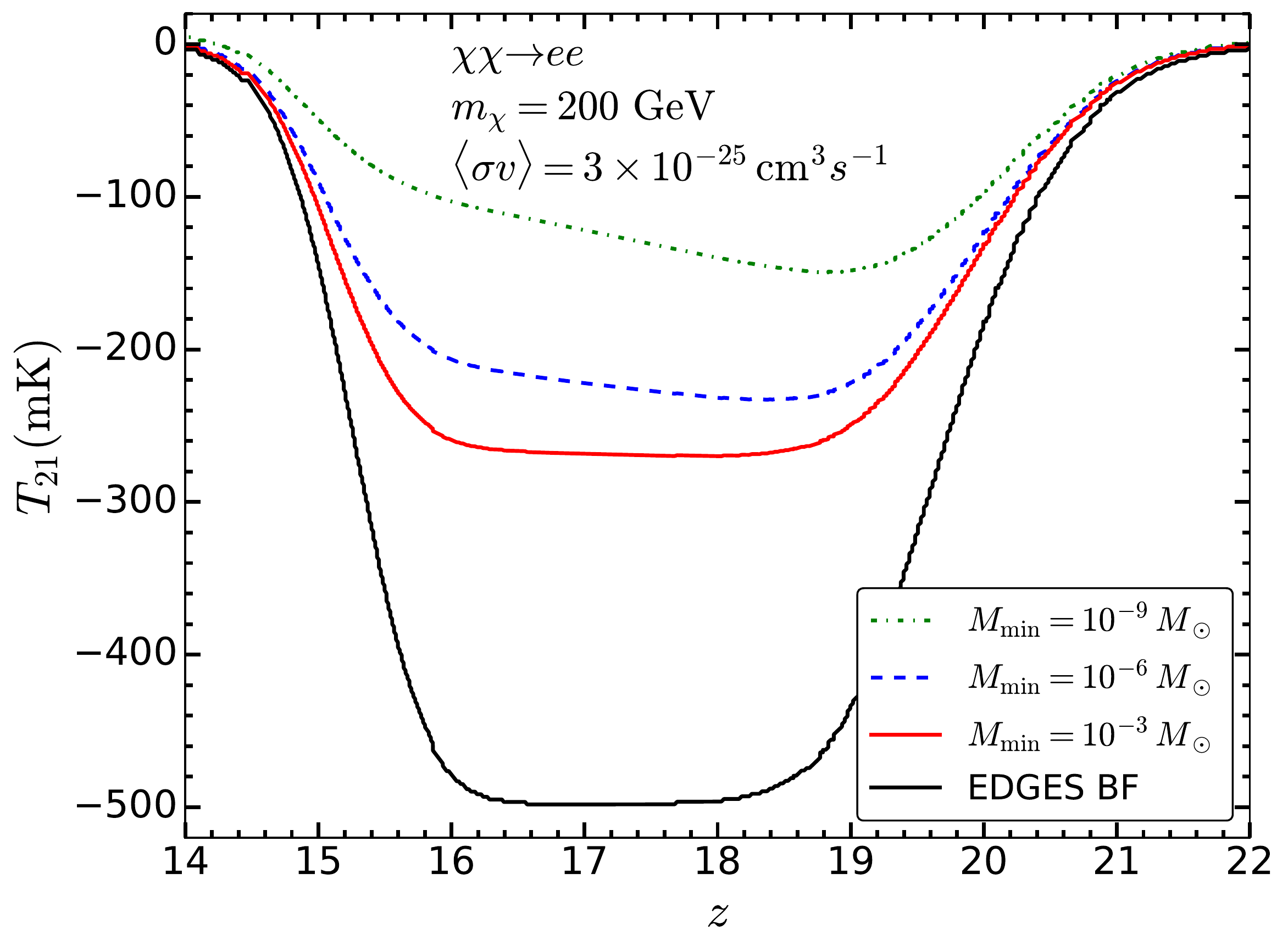}
\caption{
Evolution of $T_{21}$ with three different minimum halo masses. 
The black line is the EDGES best-fit model.
\label{fig:boost}}
\end{figure}

Similar to Ref.~\cite{DAmico:2018sxd}, we also introduce a cosmological 
boost factor $\mathcal{B}$ 
to account for the effect of DM inhomogeneities and 
structures~\cite{Evoli:2014pva}, 
\begin{align}
\label{Eq:Boost}
	 \mathcal{B}(z')=1+\frac{b_h\times\texttt{erfc}\left[(1+z')/(1+z_h)\right]}{(1+z')^{\delta}}\times 10^{5}.	 
\end{align}
In Ref.~\cite{Evoli:2014pva}, three of boost factor configurations are given; 
for the resolution of the minimum halo mass $M_{\rm{min}}=(10^{-3},~10^{-6},~10^{-9})$ solar mass, 
the parameters are $b_h=(1.6,~6.0,~23.0)$, $z_h=(19.5,~19.0,~18.6)$, and $\delta=(1.54,~1.52,~1.48)$, respectively. 
These three configurations can be treated as the representative systematic uncertainties.
In Fig.~\ref{fig:boost}, we show their impacts on the plane ($z$, $T_{21}$) for $\mchi=200\gev$ and 
$\sv=3\times 10^{-25} {\rm cm}^3 s^{-1}$ with the assumption of DM $100\%$ annihilation to electron-positron pairs. 
The black line is the EDGES best-fit model (EDGES BF).
Clearly, the red solid line with the resolution of the minimum halo mass $M_{\rm{min}}=10^{-3} M_\odot$  
gives the weakest and the most conservative limit than the other two. 
Hereafter, we will present our result only based on mass resolution $M_{\rm{min}}=10^{-3} M_\odot$ as being conservative.

\begin{figure}[!htb]
\includegraphics[width=0.8\textwidth]{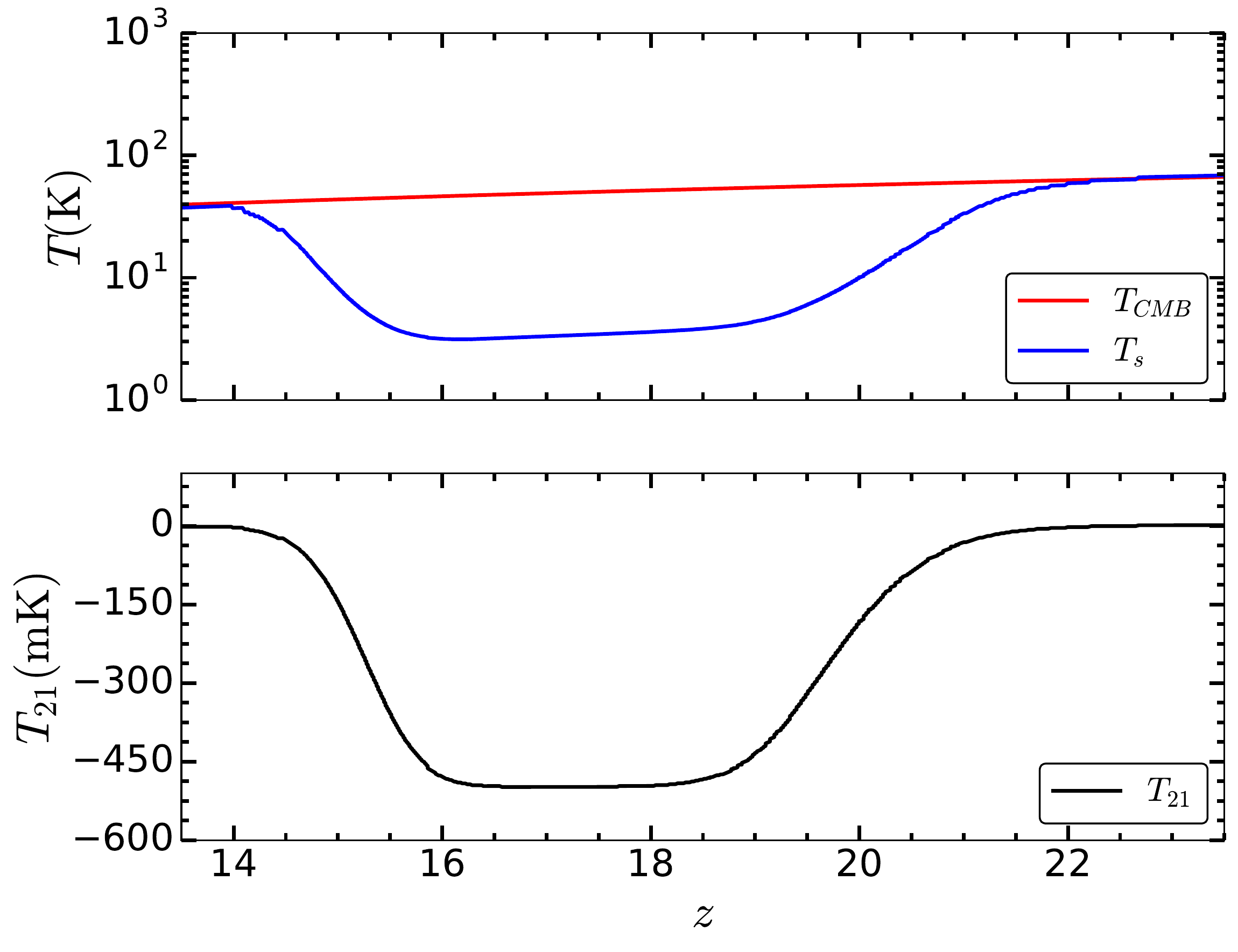}
\caption{The best-fit model for $T_{21}$ (lower panel) given in the extended data in Fig. 8 of Ref.~\cite{Bowman:2018yin}. 
The upper panel presents the $T_{s}$ which is converted from the best-fit model of $T_{21}$ by 
using Eq.~\eqref{eq:t21}.
\label{fig:Ts}}
\end{figure}

In this paragraph, we would like to explicitly demonstrate 
how we compute the theoretical prediction of the DM modified $T_{21}$ signal. 
First of all, after all the DM ingredients are included, 
we insert Eqs.~\eqref{eq:dxdz} and~\eqref{eq:dTdz} into the \texttt{RECFAST}
code~\cite{Seager:1999bc} to obtain the DM-modified matter temperature $T_{g}(z)$.
Secondly, we perform the \texttt{RECFAST} computation again by switching off the 
DM contribution in order to calculate matter temperature without DM annihilation. 
Then, comparing the matter temperature from both scenarios, 
one can obtain the change of $T_{g}$ by DM annihilation $\delta T_{g}$. 
This quantity is useful in the next step. 
Before performing the third step, let us make a reasonable assumption that $T_{s}$ is fully coupled to $T_{g}$ at
$z=15-20$ as indicated by the $T_{21}$ signal probed by EDGES.    
Therefore, in the third step, we can simply obtain 
the DM-modified spin temperature by adding the
difference $\delta T_{g}$ 
to the best-fit spin temperature $T_{s}^{\rm{BF}}$ which is converted from the EDGES best-fit $T_{21}$ signal 
shown in Fig.~\ref{fig:Ts}. 
Finally, plugging this new spin temperature $T_{s}^{\rm{BF}}+\delta T_{g}$ into Eq.~\eqref{eq:t21}, 
one can simply compute DM modified $T_{21}$ without involving any unknown 
astrophysical source other than the DM annihilation. 
In some sense, our approach can be treated as a background-free approach because  
all the known or unknown astrophysical sources are absorbed in the EDGES best-fit $T_{21}$ signal.

Assuming that DM annihilation can only contribute sub-dominantly to the EDGES signal, one can invert
Eq.~\eqref{eq:t21} to obtain $T_s$, to which DM does not contribute much.
In Fig.~\ref{fig:Ts}, we present the best-fit residual model for
$T_{21}$~\cite{Bowman:2018yin} (lower panel) after the foreground is
properly removed.  In the upper panel of Fig.~\ref{fig:Ts}, we present
the resulted $T_s$ in comparison with $T_{CMB}$.  
As such, the data-derived $T_s$ can be treated as a base model to derive upper limits on
DM annihilation.
There lacks likelihood information in Ref.~\cite{Bowman:2018yin}. However, Ref.~\cite{Barkana:2018lgd}
claimed that
the standard value predicted by the first-star model 
$T_{21}=-209\mk$
is $3.8\sigma$ away from the
experimental value $T_{21}^{\texttt{EDGES}}=-500\mk$.  By assuming a
Gaussian distribution, one can approximate the standard deviation as
$\sigma_{\texttt{EDGES}}=(500-209)/3.8\mk$ and hence the likelihood
can be rewritten as
\begin{eqnarray}
\mathcal{L} \propto \exp\left[-\frac{\chi^2}{2}\right],    {\rm ~~where~~}  
\chi^2=\frac{\left(T_{21}^{\texttt{EDGES}}-T_{21}^{\texttt{TH.}}\right)^2}{\sigma_{\texttt{EDGES}}^2}.\nonumber
\label{eq:chisq}
\end{eqnarray} 
The standard deviation captures the EDGES experimental uncertainty that limits the relative $T_{21}$. As mentioned above, the absolute level of the 21-cm absorption is irrelevant in the present consideration.

Here are some comments on the choice of our likelihood. 
Firstly, one might concern that the usage of the Gaussian likelihood would be improper 
if the analysis is with a tiny signal. 
However, one should bear in mind that the signal $T_{21}^{\texttt{TH.}}$ in Eq.~\eqref{eq:chisq} 
includes both contributions from the EDGES best-fit model and the DM annihilation, despite the nonlinearity.  
For the EDGES best-fit model, the reported signal-to-noise ratio is 37 at frequency of 78.1 MHz with the amplitude of 0.53 K 
which is strong enough to treat its likelihood as Gaussian.  
When adding the DM annihilation contribution, $T_{21}^{\texttt{TH.}}$ will only be larger because of more energy injection. 
Hence, it is reasonable to use the Gaussian likelihood. 
Secondly, under such a likelihood function, 
we directly use the experimental error bar of $T_{21}$ measured by EDGES 
to constrain DM annihilation. 
Namely, it is a \textit{background-only} likelihood  that the \textit{background} refers to everything else 
beside the contribution from WIMP DM annihilation. 
Being not to explain the excess with DM annihilation, the upper limit on the annihilation cross section 
derived from the likelihood in Eq.~\eqref{eq:chisq} is expected to be weaker 
but more conservative than the true one.

\section{Results\label{sec:app}}

\begin{figure}[!htb]
\includegraphics[width=0.7\textwidth]{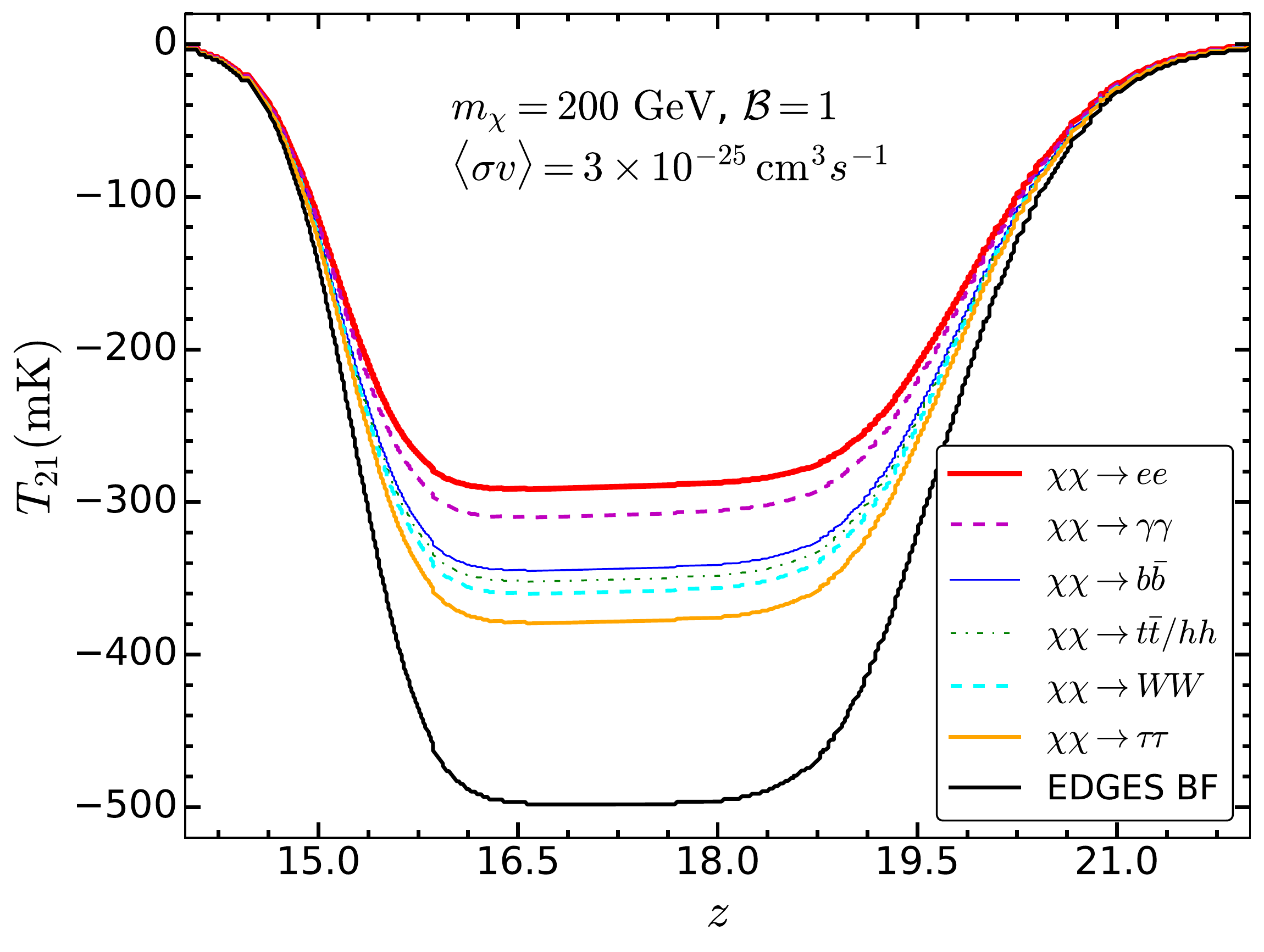}
\caption{
$T_{21}$ evolution for different DM annihilation channels with $\mchi=200\gev$.
\label{fig:DM_ch}}
\end{figure}

As aforementioned, WIMP DM has a tiny scattering cross section 
with SM nucleon. Therefore, one can safely ignore the gas cooling effect and 
put an upper limit on the DM annihilation that contributes to the 
gas heating. To consequently constrain the annihilation cross section, 
the energy spectra ($\dnede$ and $\dngde$) in Eq.~\eqref{eq:dchidz}
have to be determined first and they depend on annihilation channels. 
Different annihilation channels will result in different sizes of $T_{21}^{\rm{TH.}}$. 
In Fig.~\ref{fig:DM_ch}, we summarize the results of $T_{21}$ for different DM annihilation channels. 
We take DM mass at $200\gev$ and annihilation cross section of $\sv=3\times 10^{-25}~{\rm cm}^3 s^{-1}$ as an example. 
To explicitly show the differences between different channels, we do not include the boost factor in these results.
The EDGES best-fit data is depicted with a black solid line for comparison. 
Clearly, DM annihilation to an electron-positron pair (red solid line) is the strongest one to increase the temperature 
while the $\tau^+\tau^-$ final-state channel is the weakest one.      
Therefore, we choose the strongest lepton channel $\chi\chi\to e^+e^-$ and one of the
strongest quark channels $\chi\chi\to b\bar{b}$ as the representative cases, while other channels are weaker than these two.

\begin{table}
\centering
\begin{tabular}{|c|c|c|c|c|c|c|}
\hline 
\hline 
$\Omega_\Lambda$ & $H_0$ & $\Omega_b$ & $\Omega_b h^2$ & $\Omega_c h^2$ & $\tau$ & $n_s$ \\ 
\hline 
~0.6844~ & ~67.27~ & ~0.0492~ & ~0.02225~ & ~0.1198~ & ~0.079~ & ~0.9645~ \\ 
\hline 
\hline 
\end{tabular} 
\caption{ The cosmology parameters used in this work.
 The values are taken from the Planck 2015 TT,TE,EE+lowP central values~\cite{Ade:2015xua}.
\label{tab:cosp}}
\end{table}

To compute the background evolution, we adopt the cosmological parameters as listed in Table~\ref{tab:cosp}.
Here we ignore the inherent 
uncertainties in these parameters
because the EDGES $T_{21}$ experimental errors dominate the uncertainties
of the current analysis.

\begin{figure}[!htb]
\includegraphics[width=0.45\textwidth]{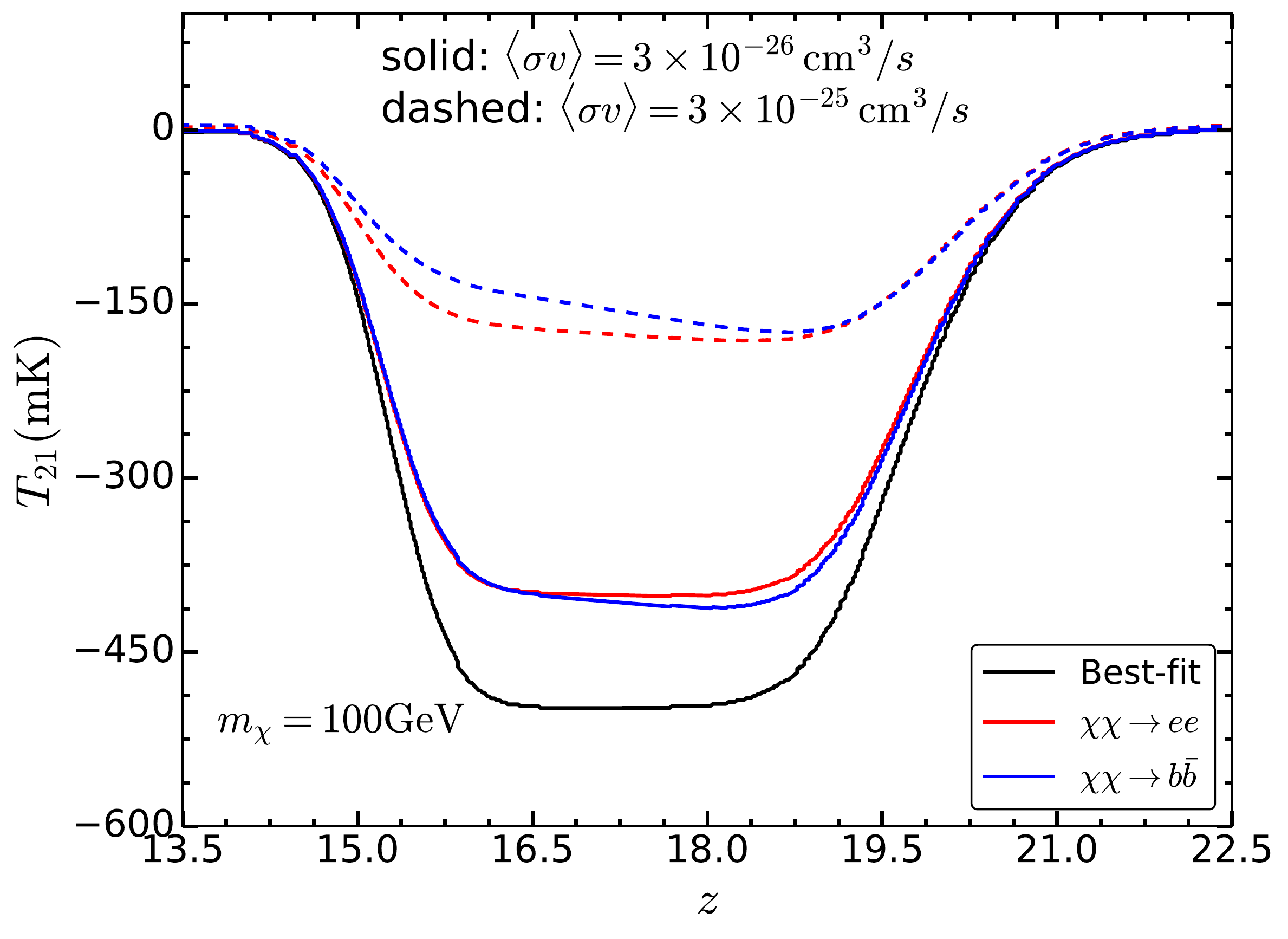}
\includegraphics[width=0.45\textwidth]{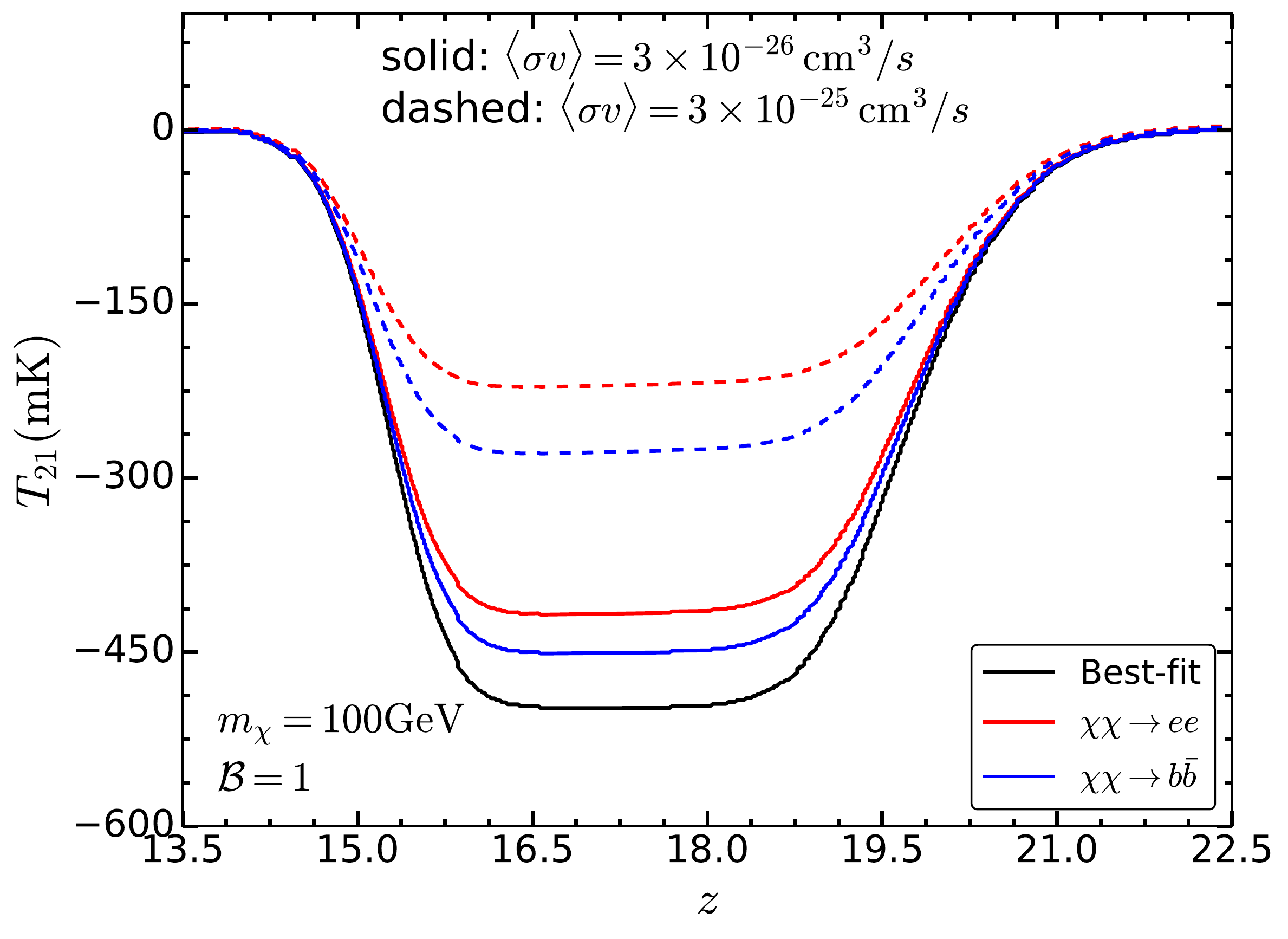}
\caption{ Comparison of the effects from different DM annihilation channels and cross sections on the $T_{21}$ signal,
with the boost factor (left panel) and without the boost factor (right panel).
The black solid line is the $T_{21}$ signal from EDGES. 
The red lines represent the modified $T_{21}$ temperatures with DM annihilation process $\chi\chi \rightarrow e^{+} e^{-}$ while the blue lines represent those with DM annihilation process $\chi\chi \rightarrow b\bar{b}$.
\label{fig:Tscf}}
\end{figure}

The effects of DM annihilation have approximately linear dependence on the 
cross sections for $z > 300$. This is a general feature of CMB constraints on DM annihilation
that affects $x_e$ in $z\sim 600-1100$~\cite{Chen:2003gz,Slatyer:2009yq,Evoli:2012zz}.     
However, during the reionization epoch ($z\sim 20$) the linearity disappears
and the behavior becomes strongly dependent on the DM property and 
the injected energy propagation~\cite{Lopez-Honorez:2016sur,Liu:2016cnk,Poulin:2016anj}. 
In the redshift region below $z=10$, where $T_{21}$ is completely governed 
by the first stars, we rely on the reionization model for $x_e$ in Ref.~\cite{Ade:2015xua}.

%
Energetic $e^+e^-$ pairs and photons can propagate a farther
distance before they are absorbed, namely, energy absorption into the gas 
at lower redshifts more likely comes from lower-energy $e^+e^-$ pairs
and photons being created by energetic particles in the gas neighborhood. 
Hence, there are two factors that cause the growth of the $b\bar{b}$ 
contribution over the $e^+e^-$ channel in heating the gas at low redshifts ($z<20$). 
First, the DM annihilation to $b\bar{b}$ can yield 
more lower-energy $e^+e^-$ pairs and photons than the $e^+e^-$ channel. 
Second, the low-energy $e^+e^-$ pairs and photons created at low redshifts 
will be shortly absorbed and successively affecting the recent epoch.
Moreover, the boost factor can largely enhance the annihilation contribution, leading to higher
 $T_g(\chi\chi\to b\bar{b})$ temperatures at low redshifts.

In Fig.~\ref{fig:Tscf}, we compare the experimental best-fit model of $T_{21}$
with those $T_{21}$ of four different DM annihilation scenarios. 
Here, we fix the DM mass to be $100\gev$ and the annihilation cross sections 
to be $3\times 10^{-26}~\rm{cm}^3 s^{-1}$ (solid line) 
and $3\times 10^{-25}~\rm{cm}^3 s^{-1}$ (dash line). 
As explained previously, $T_{21}$ at low $z$ should be more altered by
$\chi\chi\to b\bar{b}$ (blue line) than $\chi\chi\to e^+e^-$ (red line). 
When comparing the results from $\chi\chi\to b\bar{b}$ and $\chi\chi\to e^+e^-$, 
we found that the boost factor is not so effective for the $e^+e^-$ channel.

\begin{figure}[!htb]
\includegraphics[width=0.48\textwidth]{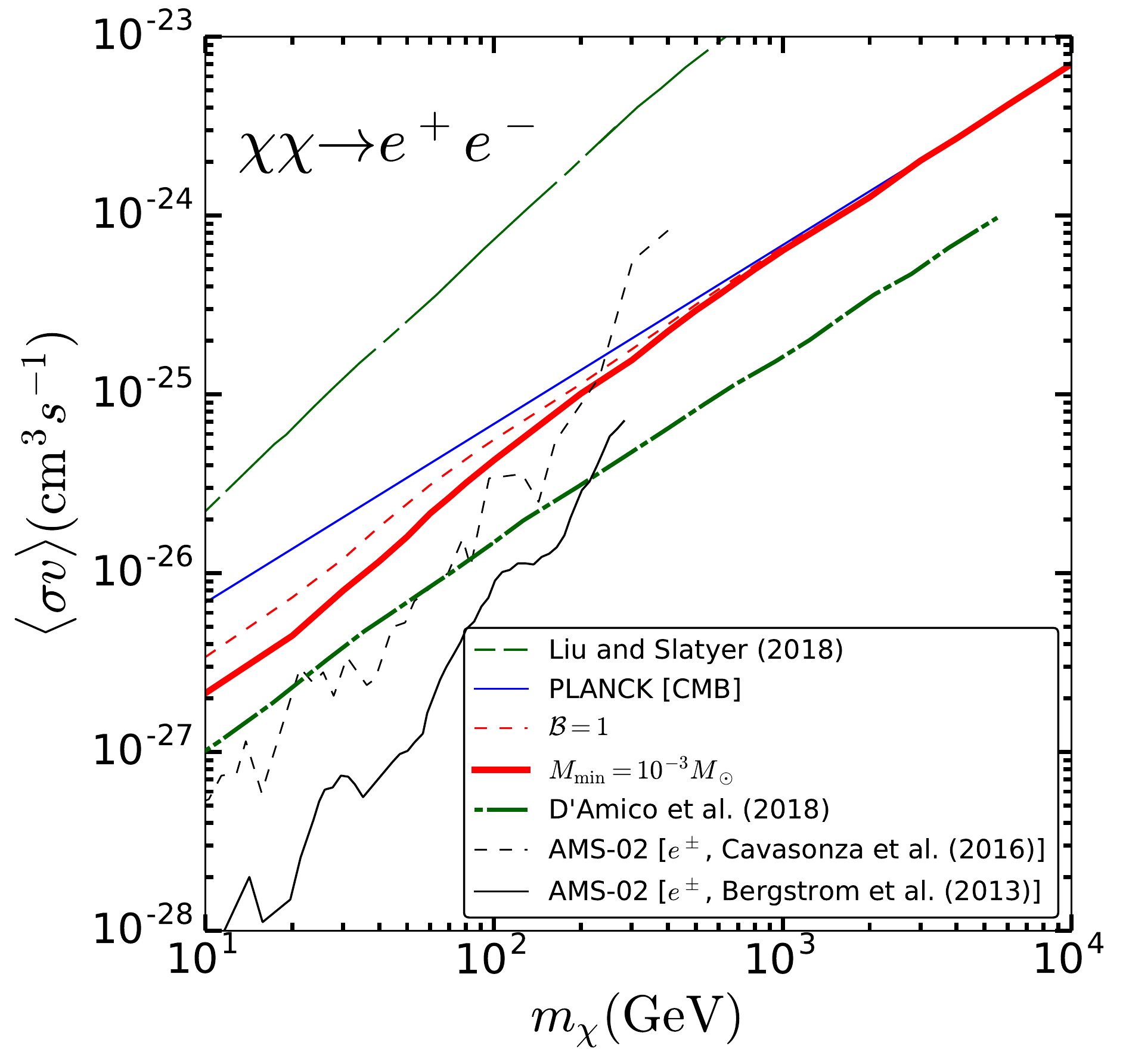}
\includegraphics[width=0.48\textwidth]{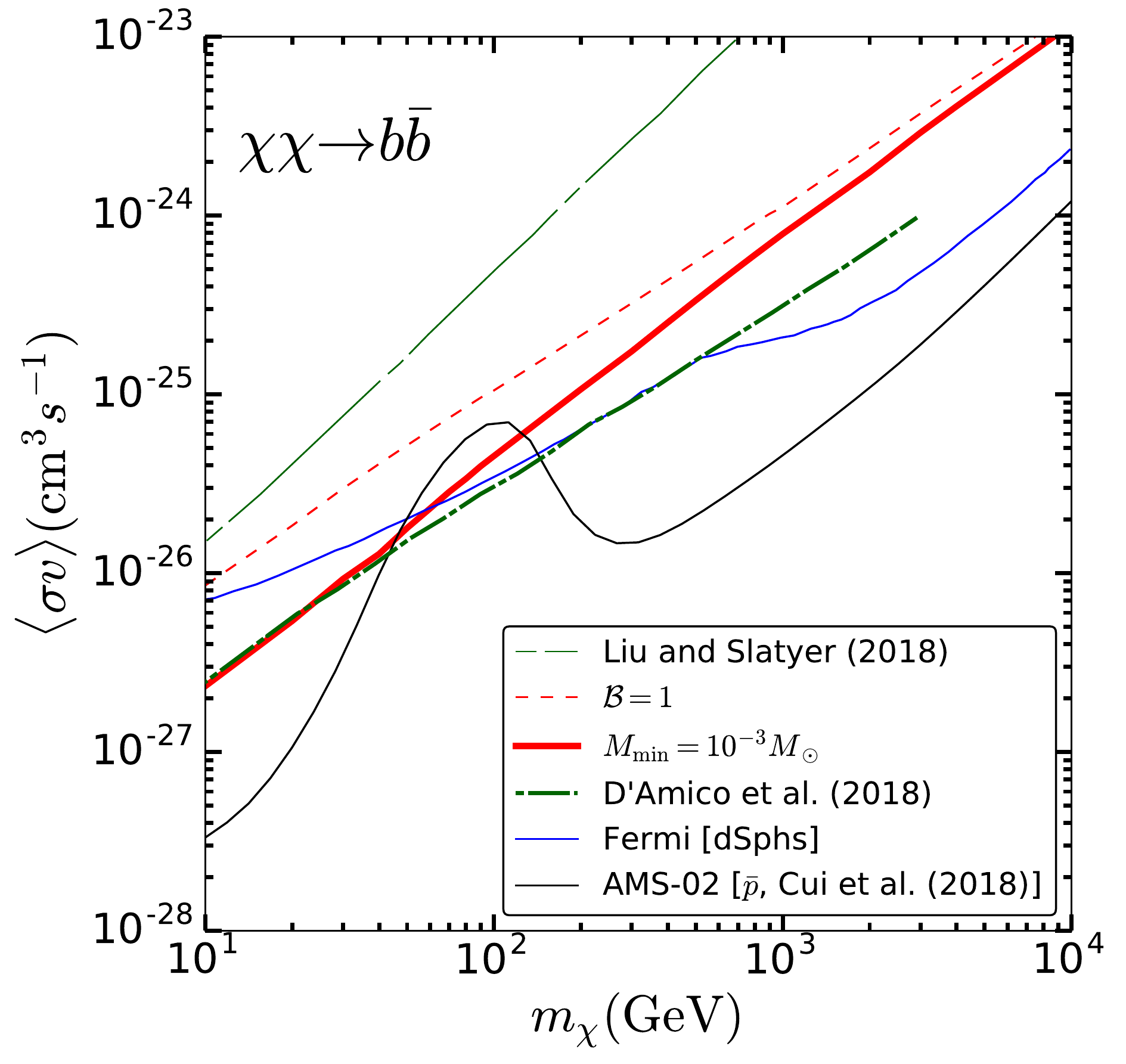}
\caption{
The 95\% upper limits on the DM annihilation cross
sections versus the DM mass derived from fitting EDGES data 
for the $e^+e^-$ channel (left panel) and the $b\bar{b}$ channel (right panel), 
denoted by the solid red lines with boost factors included
and the dashed red lines without boost factors.
The 95\% exclusion limits from PLANCK CMB~\cite{Ade:2015xua}, 
Fermi dSphs~\cite{Fermi-LAT:2016uux}, 
and AMS02 data for electron-positron~\cite{Bergstrom:2013jra,Cavasonza:2016qem} and 
antiproton~\cite{Cui:2016ppb,Cui:2018klo}
are also shown. 
For comparison, we plot two recent results from Liu and Slatyer~\cite{Liu:2018uzy} (green long dashed line)
and D'Amico {\it et al.}~\cite{DAmico:2018sxd} (green short dashed line). 
The reference limits taken from ~\cite{Liu:2018uzy} are based on the scenario: 
non-standard recombination allowing the gas to decouple thermally from the CMB earlier 
with $T_g(z=17.2) = 5.2\,\mathrm{K}$ and $(1+z)_\mathrm{td} = 500$ where ``$\mathrm{td}$" in the subscript stands for thermal decoupling.
\label{fig:constraint}}
\end{figure}

Finally, we present the $95\%$ exclusion limits in the 
($\mchi$, $\sv$) plane based on EDGES data in Fig.~\ref{fig:constraint}.
For each DM mass bin, we normalize the background-only
likelihood to one and depict a solid line in $95\%$ C.L. 
corresponding to $-2\Delta \ln \mathcal{L}=2.71$ for a one-sided 
Gaussian likelihood distribution.

With the boost factor (red solid lines), one can see that the $e^+e^-$ channel (left panel) is only 
slightly more stringent than the $b\bar{b}$ (right panel) channel in the lower-mass region 
but it becomes stronger at higher masses.
However, for the case without the boost factor (red dashed lines), 
the $e^+e^-$ channel is overall a factor of $2-3$ more stringent than 
the $b\bar{b}$ channel. 
Interestingly, the boost factor does not affect the $e^+e^-$ 
channel as significantly as the $b\bar{b}$ channel. 
This is totally due to the fact that more energetic $e^+e^-$ pairs or photons  
are absorbed in a longer distance as aforementioned. 
For $\mchi\gtrsim 500\gev$, 
one can see no difference in the $e^+e^-$ channel
between the cases with and without the boost factor.    

Both our EDGES limits for the $e^+e^-$ and $b\bar{b}$ channels are stronger
than the PLANCK CMB constraint, denoted by a blue thin line in the left panel only.
When comparing with the Fermi dSph data (blue thin line of right panel), 
the EDGES limit for the $b\bar{b}$ channel is stronger than the Fermi constraint in the 
low-mass region ($m_\chi \leq 80\,\mathrm{GeV}$) 
while being weaker at higher masses.
However, if the boost factor is removed, the EDGES limit will be no longer stronger than the
Fermi limit.
On the other hand, the cosmic ray constraints are in general
stronger than the EDGES limits. 
For the $e^+e^-$ final states, the derived limits from the AMS-02 electron-positron measurement (black solid/dashed lines of left panel) 
are more stringent than the EDGES limit at the DM mass less than around $300\gev$.
The AMS-02 antiproton constraint (black solid line of right panel) is
stronger than the EDGES limit for the $b\bar{b}$ channel
even when the boost factor is included, except for 
a small mass window ($m_\chi = 30 - 200 \,\mathrm{GeV}$).
Overall, for a thermal DM relic whose annihilation is through a
$s$-wave (velocity independent) process with a standard value of $\sv\sim 3\times 10^{-26} \,\mathrm{cm}^3 s^{-1}$, 
the EDGES limits can exclude both $e^{+}e^{-}$ and $b\bar{b}$ channels
in the mass region $\mchi\leq 100\gev$ at 95\% C.L.

We also depict the relevant results from D'Amico {\it et al.}~\cite{DAmico:2018sxd} (green short-dashed lines)
and Liu and Slatyer~\cite{Liu:2018uzy} (green long-dashed lines). 
The green short-dashed and the red solid lines are based on the same boost factor, 
whereas there is no boost factor applied in the green long-dashed and the red dashed lines.
Note that the statistics between our constraints and those 
in Refs.~\cite{DAmico:2018sxd} and ~\cite{Liu:2018uzy}
is clearly different.
The green long-dashed line is taken out from Fig.~6~(b) of Ref.~\cite{Liu:2018uzy}
with the $e^+ e^-$ channel and $b\bar b$ channel.
In Ref.~\cite{DAmico:2018sxd}, the constraints are obtained by demanding that the
standard value of the 21-cm absorption ($T_{21}=-200\,\mathrm{mK}$) is not washed 
out by the DM annihilation by half the amount ($\Delta T_{21} \leq 100\,\mathrm{mK}$). 
We note that our result is similar to that in Ref.~\cite{DAmico:2018sxd} at the lower DM mass  
because the one-sigma uncertainty in $T_{21}$ is $T_{21}^{95\%}\sim 126\,\mathrm{mK}$ in our statistical approach.   
In Ref.~\cite{Liu:2018uzy}, their upper limits are obtained under the criteria 
that DM-modified gas temperature $T_g<5.2$ K at $z=17.2$. 
On the other hand, our analyses are based
on the best-fit model of the EDGES $T_{21}$ signal 
to derive the constraints on the DM annihilation.
Our limits are slightly weaker than those in Ref.~\cite{DAmico:2018sxd} but 
stronger than Ref.~\cite{Liu:2018uzy}.

Beside the statistics, we do not assume immediate absorption of electrons, positrons, and photons during the propagation from $z'$ to $z$. 
This assumption is applied in Refs.~\cite{DAmico:2018sxd}, where it relies on fitting energy fractions $\feff$. 
On the other hand, Ref.~\cite{Liu:2018uzy} models the heating and ionization due to gradual cooling of the injected
electron/positron/photon over time.
Note that we only adopt the non-standard recombination scenario~\footnote{There are three general scenarios 
discussed in Ref.~\cite{Liu:2018uzy}:
(i) additional radiation backgrounds in the frequency range surrounding 21-cm,
(ii) non-standard recombination allowing the gas to decouple thermally from the CMB earlier, and
(iii) cooling of the gas through DM-baryon scattering. 
Our WIMP scenario can be either the scenario (i) or (ii). 
However, there is no public available $b\bar{b}$ annihilation limit for scenario (i) in Ref.~\cite{Liu:2018uzy}. 
We only present the scenario (ii) but the limits are similar, see Fig.~(12c) in Ref.~\cite{Liu:2018uzy}.     
} 
of Ref.~\cite{Liu:2018uzy} which 
the gas decouples thermally from the CMB earlier, see the right panel of Fig.~6 of Ref.~\cite{Liu:2018uzy}.

There are three factors that can make the approximation of a simple $\feff$ fraction invalid:
(1) when one includes the boost factor $\mathcal{B}(z)$, taking delayed deposition into account is particularly important;
(2) the efficiency of deposition at $z\sim 17$ is quite different than at the high redshifts ($z\sim 600$) that dominate the CMB constraints;
(3) the $\feff$ values in Refs.~\cite{Slatyer:2012yq,Slatyer:2015kla} are chosen to give equivalent energy transfers to the ionization of the gas, but not to the heating.
\footnote{We gratefully acknowledge the private communication with T.~Slatyer, 
the author of Refs.~\cite{Slatyer:2012yq,Slatyer:2015kla,Liu:2018uzy}.}
We have tried to extract redshift-dependent $\feff$ fractions from our full numerical calculations using Eqs.~\eqref{eq:dxdz} and \eqref{eq:dTdz}. Our results show that the $\feff$'s for ionization vary slowly with redshift in the range of $\feff\sim 0.1-0.4$, quantitatively consistent with those given in 
Refs.~\cite{Slatyer:2012yq,Slatyer:2015kla}, and that the heating $\feff$'s rise monotonically with redshift from $0.1$ to $0.5$. Note that the $\feff$ here has different definition with those given in 
Refs.~\cite{Slatyer:2012yq,Slatyer:2015kla}. We will report the details in the Appendix~\ref{sec:appA}.

\section{Conclusions}

Regarding the absorption signal of 21-cm line probed by the EDGES experiment, we have assumed
null DM contribution in signal to derive the modified evolution of $T_{21}$
by considering DM annihilation into $e^{+}e^{-}$ and $b\bar{b}$ final states. 
We have further assumed that the process of DM annihilation is velocity independent ($s$-wave)
and considered the boost factor as for approximating the enhancement 
of DM annihilation in the cosmic structures.
Different from other studies that simply make use of effective energy fractions $\feff$, 
we directly compute the propagation of the injected energy rather than assuming instant energy deposition.
%
%
It is precise enough to obtain the CMB constraints on DM annihilation by using $\feff$,
since the effects take place linearly at high redshifts.
However, for the 21-cm absorption at low redshifts, the linearity breaks down due
to the formation of structures, and thus the assumption of using a simple $\feff$ may not be adequate.
We have shown the effects of the boost factor on the $e^{+}e^{-}$ and $b\bar{b}$ channels and 
found that the enhancement is larger for the latter due to 
the fact that the $b\bar{b}$ final states produce more low-energy $e^{+}e^{-}$ pairs and photons
after a sequence of cascading decays.
As expected, our EDGES limit for the $e^{+}e^{-}$ channel is more stringent than that for the $b\bar{b}$ 
channel. 
Nevertheless, with the enhancement of the boost factor, the two channels are almost tantamount.
Overall, we have derived the 21-cm constraints on the DM annihilation to $e^{+}e^{-}$ and 
$b\bar{b}$ channels, which exceed the CMB constraints and in some low-mass regions better than the limits from the Fermi dSphs data and the AMS-02 antiproton data. 
%

\acknowledgments

We thank M. Kawasaki, K. Nakayama, and T. Sekiguchi for providing us the tables.
Y.S.T. would like to thank K. Nakayama for kindly explaining
their tables and many email exchanges. This work was supported in part
by the Ministry of Science and Technology, Taiwan, ROC under Grants
No. MOST-104-2112-M-001-039-MY3 (K.W.N.) and MOST-105-2112-M-007-028-MY3 (K.C.).

 
\appendix

\section{The effective $\feff$. }
\label{sec:appA}
 
\begin{figure}
\begin{center}
\includegraphics[width=0.45\textwidth]{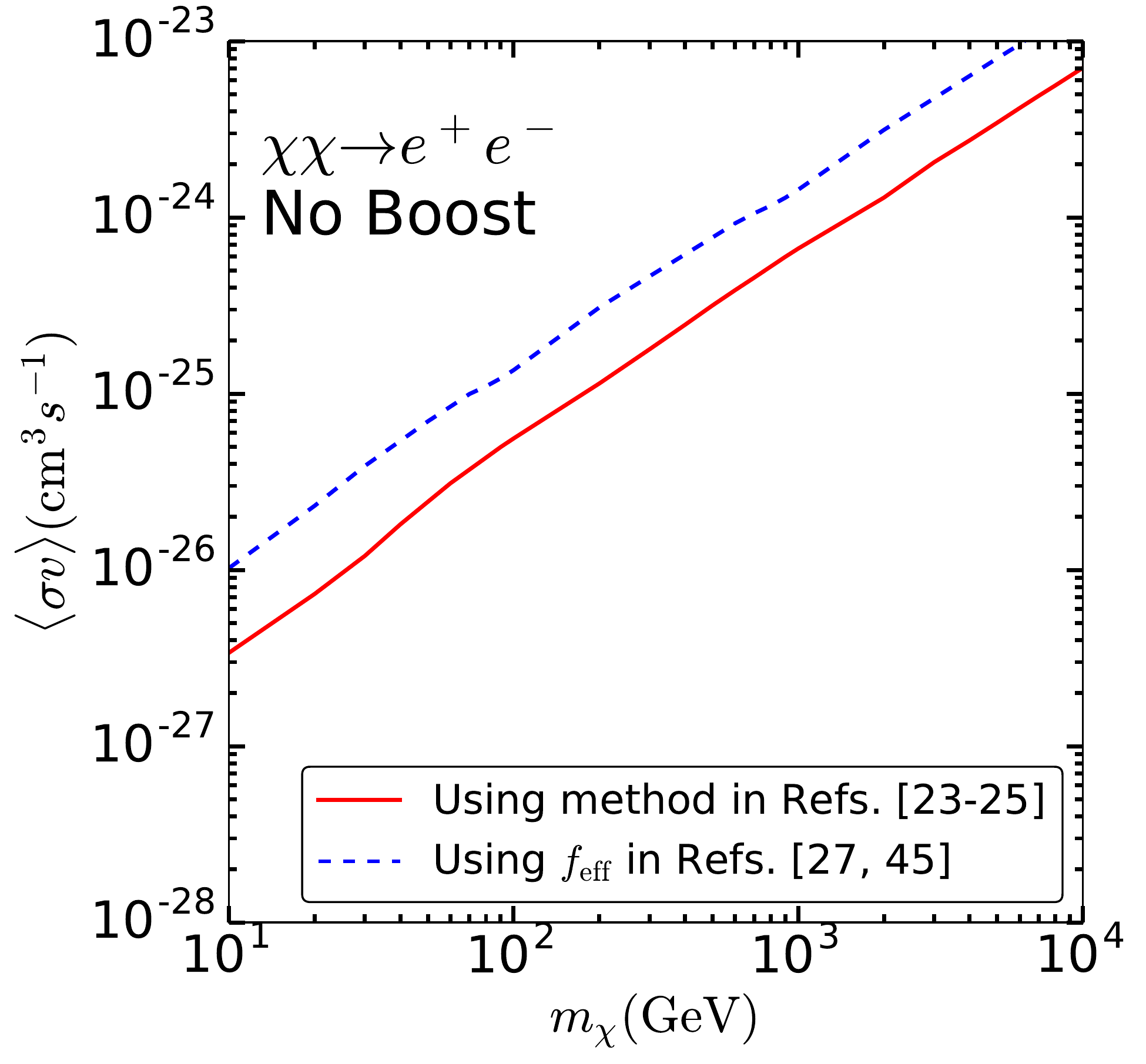}
\includegraphics[width=0.45\textwidth]{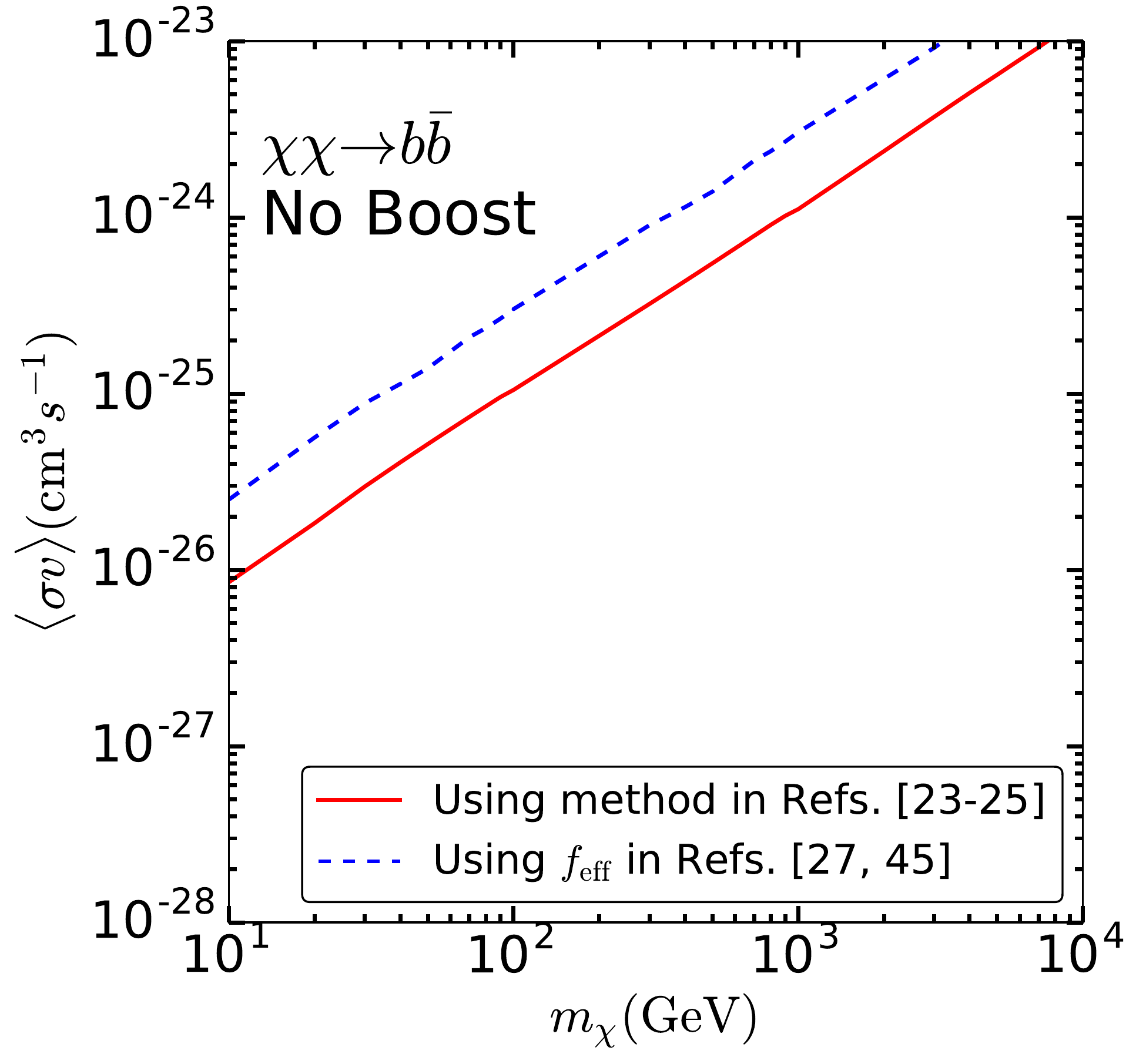}
\caption{Comparison between two different approaches. 
There is no boost factor applied for $e^+e^-$ final states (left) and 
$b\bar{b}$ final states (right).
\label{Fig:cf_feff}}
\end{center} 
\end{figure}

In Fig.~\ref{Fig:cf_feff}, using the same EDGES likelihood, 
we compare the formalism and methodology developed in 
Refs.~\cite{Kanzaki:2008qb,Kanzaki:2009hf,Kawasaki:2015peu} (red solid lines) and 
the $\feff$ method developed in Refs.~\cite{Slatyer:2015kla,Slatyer:2015jla} (blue dashed lines). 
For the $\feff$ method we assume instantaneous energy deposition in the calculation and adopt the ``SSCK approximation"~\cite{Shull:1982zz,Chen:2003gz} which is also adopted in Ref.~\cite{DAmico:2018sxd}:
\begin{equation}
f_{\texttt{eff},i} = \feff\dfrac{1-x_e}{3},\quad f_{\texttt{eff},h} = \feff\dfrac{1+2x_e}{3}.
\end{equation}
We use the redshift and channel-independent (for ionization or heating) $\feff$ given in Ref.~\cite{Slatyer:2015jla}
for different injected energy (DM mass) and annihilation channel.
We found that the former method used in this paper is about a factor of $3$ in both channels
better than the $\feff$ method. However, if including the boost factor, 
such a difference will be reduced as shown in Fig.~\ref{fig:constraint}.

\begin{figure}
\begin{center}
\includegraphics[width=0.45\textwidth]{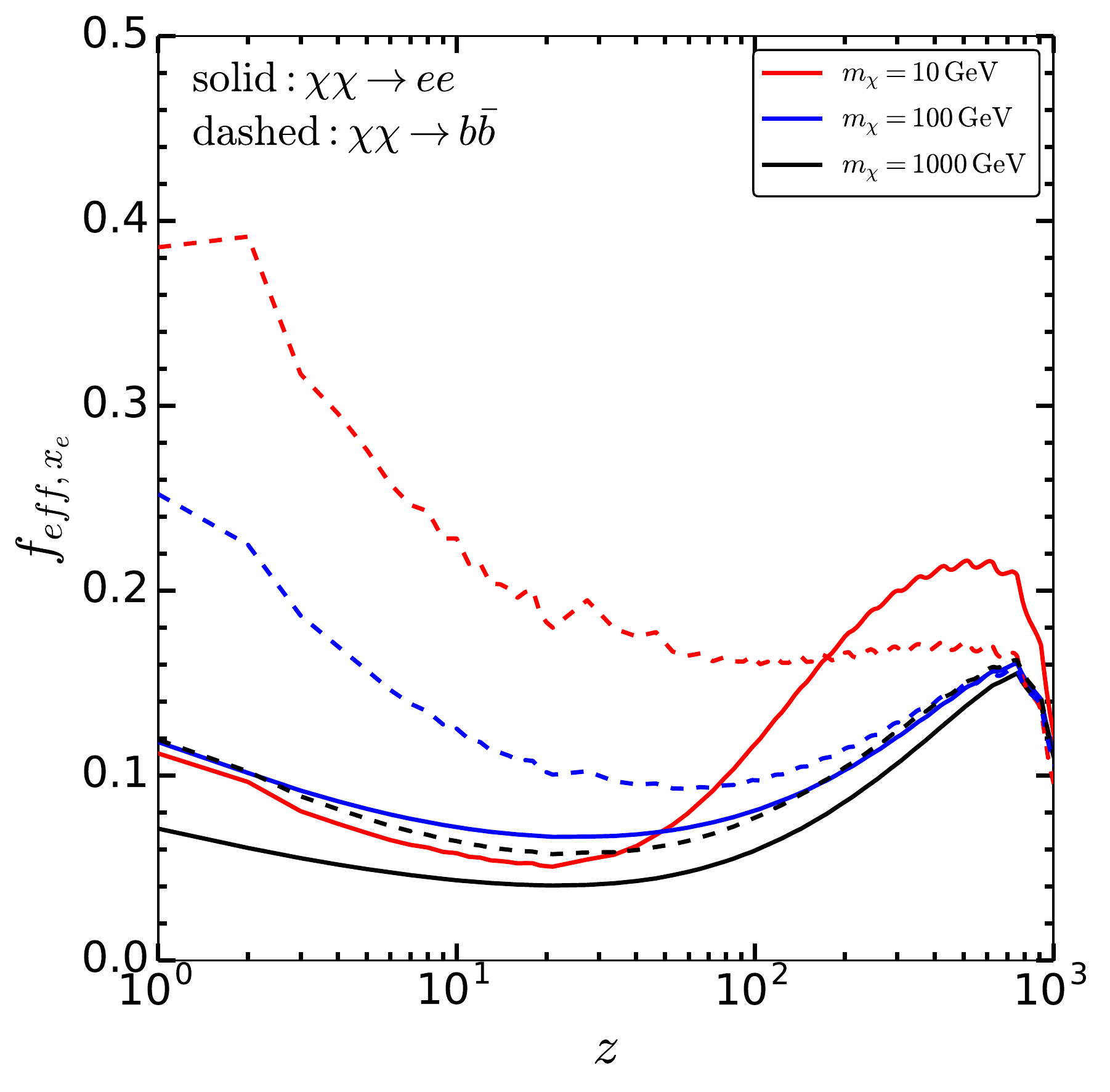}
\includegraphics[width=0.45\textwidth]{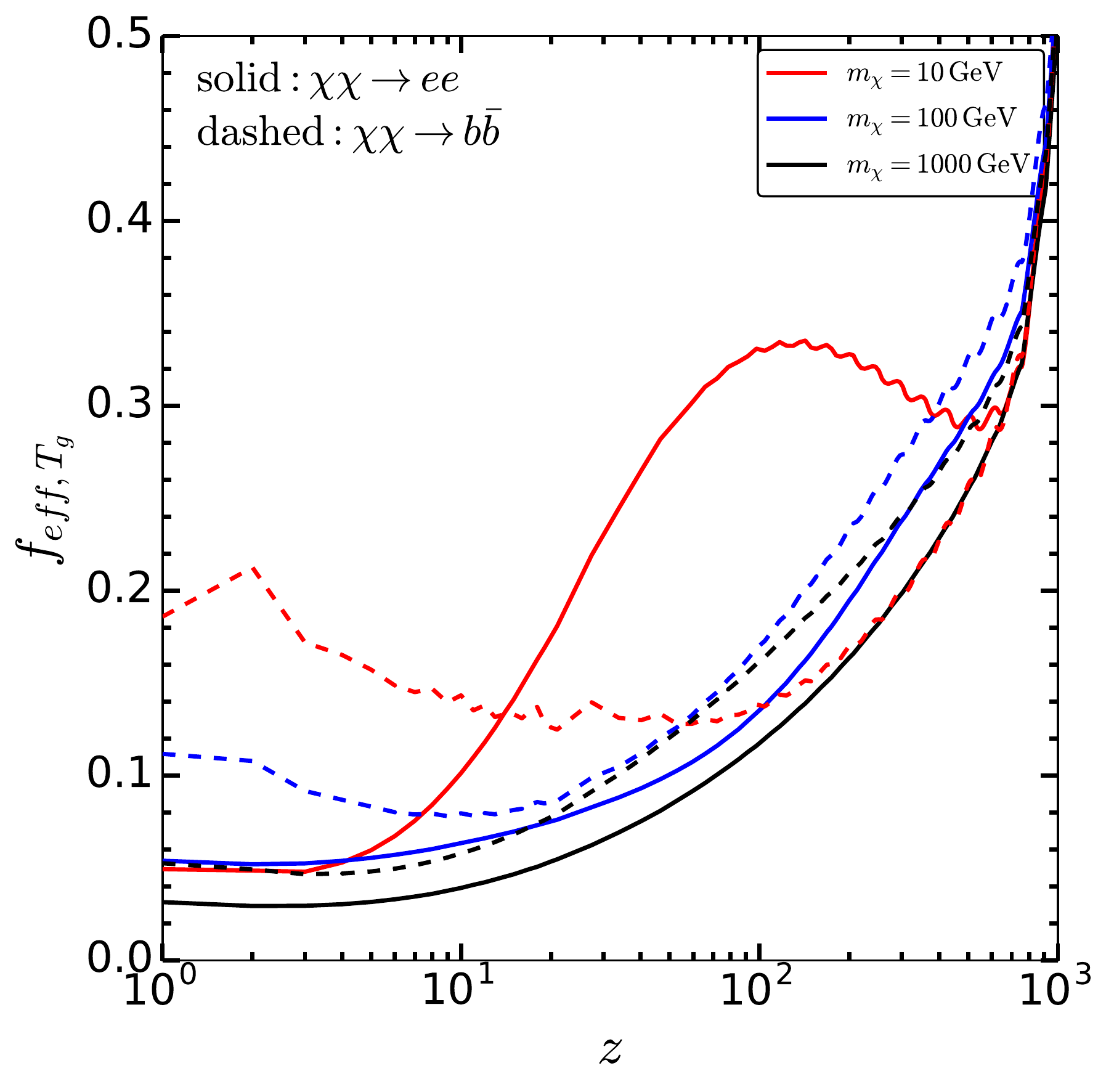}
\caption{Left panel: the $\feff$ for ionization in the processes $\chi\chi \rightarrow e^+ e^-$ and $\chi\chi \rightarrow b\bar{b}$. Right panel: the $f_{eff}$ for heating in the processes $\chi\chi \rightarrow e^+ e^-$ and $\chi\chi \rightarrow b\bar{b}$.\label{Fig:feff}}
\end{center}
\end{figure}

For further discussion of the difference, we also compute the $\feff$ based on our methodology. 
The result is shown in Fig.~\ref{Fig:feff} and basically reflects that the $bb$ channel is more effective than the $ee$ channel at low redshift, 
which is also mentioned in our paper. 
Note that these are clearly different from the original $\feff$ adopted by others.
One from $T_{g}$ and the other one from $x_{e}$ calculations and they can be different. 
In the following, we will demonstrate how we extract the values of $\feff$ for ionization and heating.

The original equations for ionization fraction and gas temperature with $100\%$ DM annihilation 
branching ratio to a fixed channel $\texttt{ch},$ are
\begin{equation}
	-\left[\dfrac{dx_e}{dz}\right]_{\rm DM} =  \int_z \dfrac{dz'}{H(z')(1+z')}
	\dfrac{n_\chi^2 (z') }{2n_H (z')} \sv \mathcal{B}(z') 
	\dfrac{m_\chi}{E_{\rm Ry}} \dfrac{d\chi_i(\texttt{ch},m_\chi,z',z)}{dz},  
\end{equation}
and 
\begin{equation}
-\left[ \frac{dT_g}{dz}\right]_{\rm DM} = \int_z \frac{dz'}{H(z')(1+z')}
	\frac{n_\chi^2(z')}{3n_H(z')}\sv \mathcal{B}(z') 
	m_\chi \frac{d\chi_h(\texttt{ch},m_\chi,z',z)}{dz},
\end{equation}
where
\begin{eqnarray}
	&&\frac{d\chi_{i,h}(\texttt{ch},m_\chi,z',z)}{dz}= \nonumber\\
	&&\int dE \frac{E}{m_\chi}\left[ 
		2\frac{dN_{e}(\texttt{ch},\mchi)}{dE} \frac{d\chi_{i,h}^{(e)}(E,z',z)}{dz} + 
		\frac{dN_{\gamma}(\texttt{ch},\mchi)}{dE} \frac{d\chi_{i,h}^{(\gamma)}(E,z',z)}{dz} 
	\right].
\end{eqnarray}

Following the most popular convention, one can assume that the injected energy 
is absorbed by the gas immediately, i.e. $z = z'$. 
Since there is no propagation of injected energy, 
we can set $d\chi_{i,h}^{(\texttt{ch})}(E,z',z)/dz =1$ and 
remove the integration with respect to the redshift.
Therefore, we can further write the evolution question as 
\begin{equation}
	-\left[\dfrac{dx_e}{dz}\right]_{\rm DM,eff} = f_{\texttt{eff},i} (z,m_\chi,\sv) \lbrace    \dfrac{1}{H(z)(1+z)}
	\dfrac{n_\chi^2 (z) }{2n_H (z)} \sv \mathcal{B}(z) 
	\dfrac{m_\chi}{E_{\rm Ry}} \dfrac{d\chi'_i(\texttt{ch},m_\chi)}{dz} \rbrace,  
\end{equation}
and 
\begin{equation}
-\left[ \frac{dT_g}{dz}\right]_{\rm DM,eff} =  f_{\texttt{eff},h} (z,m_\chi,\sv) \lbrace    \frac{1}{H(z)(1+z)}
	\frac{n_\chi^2(z)}{3n_H(z)}\sv \mathcal{B}(z) 
	m_\chi \frac{d\chi'_h(\texttt{ch},m_\chi)}{dz} \rbrace,
\end{equation}
where
\begin{equation}
	\frac{d\chi'_{i,h}(\texttt{ch},m_\chi)}{dz}= 
	\int dE \frac{E}{m_\chi}\left[ 
		2\frac{dN_{e}(\texttt{ch},\mchi)}{dE} + 
		\frac{dN_{\gamma}(\texttt{ch},\mchi)}{dE}
	\right].
\end{equation}  
The $f_{\texttt{eff},i} (z,m_\chi)$ and $f_{\texttt{eff},h} (z,m_\chi)$ are the parameters we defined to represent the effective energy transfer and propagation. The values of them can be obtained through
\begin{equation}
f_{\texttt{eff},i} (z,m_\chi) = -\left[\dfrac{dx_e}{dz}\right]_{\rm DM} /  \lbrace    \dfrac{1}{H(z)(1+z)}
	\dfrac{n_\chi^2 (z) }{2n_H (z)} \sv \mathcal{B}(z) 
	\dfrac{m_\chi}{E_{\rm Ry}} \dfrac{d\chi'_i(\texttt{ch},m_\chi)}{dz} \rbrace
\end{equation}
and
\begin{equation}
f_{\texttt{eff},h} (z,m_\chi) = -\left[\dfrac{dT_g}{dz}\right]_{\rm DM} /  \lbrace    \frac{1}{H(z)(1+z)}
	\frac{n_\chi^2(z)}{3n_H(z)}\sv \mathcal{B}(z) 
	m_\chi \frac{d\chi'_h(\texttt{ch},m_\chi)}{dz} \rbrace.
\end{equation}
Note that $\feff$ is independent of the the velocity-averaged cross section $\sv$, since $\sv$ is a constant in the original integration with respect to the redshift $z'$.


\begin{thebibliography}{15} 
 
\bibitem{Boggess:1992xla} 
  N.~W.~Boggess {\it et al.},
  Astrophys.\ J.\  {\bf 397}, 420 (1992).
  doi:10.1086/171797


\bibitem{Hinshaw:2012aka} 
  G.~Hinshaw {\it et al.} [WMAP Collaboration],
  Astrophys.\ J.\ Suppl.\  {\bf 208}, 19 (2013)
  doi:10.1088/0067-0049/208/2/19
  [arXiv:1212.5226 [astro-ph.CO]].


\bibitem{Ade:2015xua} 
  P.~A.~R.~Ade {\it et al.} [Planck Collaboration],
  Astron.\ Astrophys.\  {\bf 594}, A13 (2016)
  doi:10.1051/0004-6361/201525830
  [arXiv:1502.01589 [astro-ph.CO]].


\bibitem{Gunn:1965hd} 
  J.~E.~Gunn and B.~A.~Peterson,
  Astrophys.\ J.\  {\bf 142}, 1633 (1965).
  doi:10.1086/148444



\bibitem{Pritchard:2011xb} 
  J.~R.~Pritchard and A.~Loeb,
  Rept.\ Prog.\ Phys.\  {\bf 75}, 086901 (2012)
  doi:10.1088/0034-4885/75/8/086901
  [arXiv:1109.6012 [astro-ph.CO]].


\bibitem{vanHaarlem:2013dsa}
  M.~P.~van Haarlem {\it et al.},
  Astron.\ Astrophys.\  {\bf 556} (2013) A2
  doi:10.1051/0004-6361/201220873
  [arXiv:1305.3550 [astro-ph.IM]].


\bibitem{Tingay:2012ps} 
  S.~J.~Tingay {\it et al.},
  Publ.\ Astron.\ Soc.\ Austral.\  {\bf 30}, 7 (2013)
  doi:10.1017/pasa.2012.007
  [arXiv:1206.6945 [astro-ph.IM]].




\bibitem{Maartens:2015mra} 
  R.~Maartens {\it et al.} [SKA Cosmology SWG Collaboration],
  PoS AASKA {\bf 14}, 016 (2015)
  doi:10.22323/1.215.0016
  [arXiv:1501.04076 [astro-ph.CO]].





\bibitem{Bowman:2018yin} 
  J.~D.~Bowman, A.~E.~E.~Rogers, R.~A.~Monsalve, T.~J.~Mozdzen and N.~Mahesh,
  Nature {\bf 555}, no. 7694, 67 (2018).
  doi:10.1038/nature25792


\bibitem{Barkana:2018lgd} 
  R.~Barkana,
  Nature {\bf 555}, no. 7694, 71 (2018)
  doi:10.1038/nature25791
  [arXiv:1803.06698 [astro-ph.CO]].


\bibitem{Munoz:2018pzp} 
  J.~B.~Munoz and A.~Loeb,
  arXiv:1802.10094 [astro-ph.CO].


\bibitem{Barkana:2018qrx} 
  R.~Barkana, N.~J.~Outmezguine, D.~Redigolo and T.~Volansky,
  arXiv:1803.03091 [hep-ph].


\bibitem{Berlin:2018sjs} 
  A.~Berlin, D.~Hooper, G.~Krnjaic and S.~D.~McDermott,
  arXiv:1803.02804 [hep-ph].


\bibitem{Fialkov:2018xre} 
  A.~Fialkov, R.~Barkana and A.~Cohen,
  arXiv:1802.10577 [astro-ph.CO].


\bibitem{Fraser:2018acy} 
  S.~Fraser {\it et al.},
  arXiv:1803.03245 [hep-ph].


\bibitem{Feng:2018rje} 
  C.~Feng and G.~Holder,
  Astrophys.\ J.\  {\bf 858}, no. 2, L17 (2018)
  doi:10.3847/2041-8213/aac0fe
  [arXiv:1802.07432 [astro-ph.CO]].


\bibitem{Ewall-Wice:2018bzf} 
  A.~Ewall-Wice, T.~C.~Chang, J.~Lazio, O.~Dore, M.~Seiffert and R.~A.~Monsalve,
  arXiv:1803.01815 [astro-ph.CO].


\bibitem{Mirocha:2018cih} 
  J.~Mirocha and S.~R.~Furlanetto,
  arXiv:1803.03272 [astro-ph.GA].


\bibitem{Pospelov:2018kdh} 
  M.~Pospelov, J.~Pradler, J.~T.~Ruderman and A.~Urbano,
  arXiv:1803.07048 [hep-ph].



\bibitem{Aprile:2018dbl} 
  E.~Aprile {\it et al.} [XENON Collaboration],
  arXiv:1805.12562 [astro-ph.CO].


\bibitem{Cui:2017nnn} 
  X.~Cui {\it et al.} [PandaX-II Collaboration],
  Phys.\ Rev.\ Lett.\  {\bf 119}, no. 18, 181302 (2017)
  doi:10.1103/PhysRevLett.119.181302
  [arXiv:1708.06917 [astro-ph.CO]].
  
  

\bibitem{Dowell:2018mdb} 
  J.~Dowell and G.~B.~Taylor,
  Astrophys.\ J.\  {\bf 858}, no. 1, L9 (2018)
  doi:10.3847/2041-8213/aabf86
  [arXiv:1804.08581 [astro-ph.CO]].


\bibitem{Kanzaki:2008qb} 
  T.~Kanzaki and M.~Kawasaki,
  Phys.\ Rev.\ D {\bf 78}, 103004 (2008)
  doi:10.1103/PhysRevD.78.103004
  [arXiv:0805.3969 [astro-ph]].


\bibitem{Kanzaki:2009hf} 
  T.~Kanzaki, M.~Kawasaki and K.~Nakayama,
  Prog.\ Theor.\ Phys.\  {\bf 123}, 853 (2010)
  doi:10.1143/PTP.123.853
  [arXiv:0907.3985 [astro-ph.CO]].


\bibitem{Kawasaki:2015peu} 
  M.~Kawasaki, K.~Nakayama and T.~Sekiguchi,
  Phys.\ Lett.\ B {\bf 756}, 212 (2016)
  doi:10.1016/j.physletb.2016.03.005
  [arXiv:1512.08015 [astro-ph.CO]].


\bibitem{Slatyer:2012yq} 
  T.~R.~Slatyer,
  Phys.\ Rev.\ D {\bf 87}, no. 12, 123513 (2013)
  doi:10.1103/PhysRevD.87.123513
  [arXiv:1211.0283 [astro-ph.CO]].


\bibitem{Slatyer:2015kla} 
  T.~R.~Slatyer,
  Phys.\ Rev.\ D {\bf 93}, no. 2, 023521 (2016)
  doi:10.1103/PhysRevD.93.023521
  [arXiv:1506.03812 [astro-ph.CO]].


\bibitem{Huang:2016pxg} 
  X.~Huang, Y.~L.~S.~Tsai and Q.~Yuan,
  Comput.\ Phys.\ Commun.\  {\bf 213}, 252 (2017)
  doi:10.1016/j.cpc.2016.12.015
  [arXiv:1603.07119 [hep-ph]].


\bibitem{Cirelli:2010xx} 
  M.~Cirelli {\it et al.},
  JCAP {\bf 1103}, 051 (2011)
  Erratum: [JCAP {\bf 1210}, E01 (2012)]
  doi:10.1088/1475-7516/2012/10/E01, 10.1088/1475-7516/2011/03/051
  [arXiv:1012.4515 [hep-ph]].


\bibitem{DAmico:2018sxd} 
  G.~D'Amico, P.~Panci and A.~Strumia,
  Phys.\ Rev.\ Lett.\  {\bf 121}, no. 1, 011103 (2018)
  doi:10.1103/PhysRevLett.121.011103
  [arXiv:1803.03629 [astro-ph.CO]].


\bibitem{Evoli:2014pva} 
  C.~Evoli, A.~Mesinger and A.~Ferrara,
  JCAP {\bf 1411}, no. 11, 024 (2014)
  doi:10.1088/1475-7516/2014/11/024
  [arXiv:1408.1109 [astro-ph.HE]].


\bibitem{Seager:1999bc} 
  S.~Seager, D.~D.~Sasselov and D.~Scott,
  Astrophys.\ J.\  {\bf 523}, L1 (1999)
  doi:10.1086/312250
  [astro-ph/9909275].


\bibitem{Chen:2003gz} 
  X.~L.~Chen and M.~Kamionkowski,
  Phys.\ Rev.\ D {\bf 70}, 043502 (2004)
  doi:10.1103/PhysRevD.70.043502
  [astro-ph/0310473].


\bibitem{Slatyer:2009yq} 
  T.~R.~Slatyer, N.~Padmanabhan and D.~P.~Finkbeiner,
  Phys.\ Rev.\ D {\bf 80}, 043526 (2009)
  doi:10.1103/PhysRevD.80.043526
  [arXiv:0906.1197 [astro-ph.CO]].


\bibitem{Evoli:2012zz} 
  C.~Evoli, M.~Valdes, A.~Ferrara and N.~Yoshida,
  Mon.\ Not.\ Roy.\ Astron.\ Soc.\  {\bf 422}, 420 (2012).
  doi:10.1111/j.1365-2966.2012.20624.x


\bibitem{Lopez-Honorez:2016sur} 
  L.~Lopez-Honorez, O.~Mena, A.~Moline, S.~Palomares-Ruiz and A.~C.~Vincent,
  JCAP {\bf 1608}, no. 08, 004 (2016)
  doi:10.1088/1475-7516/2016/08/004
  [arXiv:1603.06795 [astro-ph.CO]].


\bibitem{Liu:2016cnk} 
  H.~Liu, T.~R.~Slatyer and J.~Zavala,
  Phys.\ Rev.\ D {\bf 94}, no. 6, 063507 (2016)
  doi:10.1103/PhysRevD.94.063507
  [arXiv:1604.02457 [astro-ph.CO]].


\bibitem{Poulin:2016anj} 
  V.~Poulin, J.~Lesgourgues and P.~D.~Serpico,
  JCAP {\bf 1703}, no. 03, 043 (2017)
  doi:10.1088/1475-7516/2017/03/043
  [arXiv:1610.10051 [astro-ph.CO]].


\bibitem{Fermi-LAT:2016uux} 
  A.~Albert {\it et al.} [Fermi-LAT and DES Collaborations],
  Astrophys.\ J.\  {\bf 834}, no. 2, 110 (2017)
  doi:10.3847/1538-4357/834/2/110
  [arXiv:1611.03184 [astro-ph.HE]].


\bibitem{Bergstrom:2013jra} 
  L.~Bergstrom, T.~Bringmann, I.~Cholis, D.~Hooper and C.~Weniger,
  Phys.\ Rev.\ Lett.\  {\bf 111}, 171101 (2013)
  doi:10.1103/PhysRevLett.111.171101
  [arXiv:1306.3983 [astro-ph.HE]].
    
 
\bibitem{Cavasonza:2016qem} 
  L.~A.~Cavasonza, H.~Gast, M.~Kramer, M.~Pellen and S.~Schael,
  Astrophys.\ J.\  {\bf 839}, no. 1, 36 (2017)
  doi:10.3847/1538-4357/aa624d
  [arXiv:1612.06634 [hep-ph]].
  


\bibitem{Cui:2016ppb} 
  M.~Y.~Cui, Q.~Yuan, Y.~L.~S.~Tsai and Y.~Z.~Fan,
  Phys.\ Rev.\ Lett.\  {\bf 118}, no. 19, 191101 (2017)
  doi:10.1103/PhysRevLett.118.191101
  [arXiv:1610.03840 [astro-ph.HE]].


\bibitem{Cui:2018klo} 
  M.~Y.~Cui, X.~Pan, Q.~Yuan, Y.~Z.~Fan and H.~S.~Zong,
  arXiv:1803.02163 [astro-ph.HE].
  
  
\bibitem{Liu:2018uzy} 
  H.~Liu and T.~R.~Slatyer,
  Phys.\ Rev.\ D {\bf 98}, no. 2, 023501 (2018)
  doi:10.1103/PhysRevD.98.023501
  [arXiv:1803.09739 [astro-ph.CO]].
    
    
\bibitem{Slatyer:2015jla} 
  T.~R.~Slatyer,
  Phys.\ Rev.\ D {\bf 93}, no. 2, 023527 (2016)
  doi:10.1103/PhysRevD.93.023527
  [arXiv:1506.03811 [hep-ph]].
 
\bibitem{Shull:1982zz}
      Shull~J.~M. and van~Steenberg~M.,
      Astrophys. J. Suppl."
      doi:10.1086/190769
     
\bibitem{Chen:2003gz}
      Chen~Xue-Lei and Kamionkowski~Marc,
      Phys.\ Rev.\ D {\bf 70}, 043502 (2004)
      doi:"10.1103/PhysRevD.70.043502"
     [arXiv:0310473 [astro-ph]].,



  
  
\end{thebibliography}
\end{document}